\documentclass[useAMS,usenatbib,usegraphicx]{mn2e}

\usepackage{amsmath,amssymb,bm,upgreek,framed,color,rotating}
\usepackage[latin1]{inputenc}
\usepackage{JournalNames}

\def\defby{\equiv}

\begin{document}
\title[Global Structure of Protoplanetary Discs]
{Global Structure of Magnetorotationally Turbulent Protoplanetary Discs}
\author[M. Flaig et al.]{M. Flaig,$^1$ Patrick Ruoff$^{2}$, W. Kley$^1$ and R. Kissmann$^{3}$ \\
$^1$ Institut f\"ur Astronomie \& Astrophysik, Universit\"at T\"ubingen, 
Auf der Morgenstelle 10, 72076 T\"ubingen, Germany \\
$^2$ Karlsruhe Institute of Technology, Institut f\"ur Anthropomatik,
Adenauerring 2, 76131 Karlsruhe, Germany\\
$^3$ Institut f\"ur Astro- und Teilchenphysik, Universit\"at Innsbruck,
Technikerstra\ss{}e 25/8, 6020 Innsbruck, Austria}
\date{Accepted ??. Received ??; in original form ??}
\pagerange{\pageref{firstpage}--\pageref{lastpage}} \pubyear{????}
\maketitle
\label{firstpage}
\begin{abstract}
The aim of the present paper is to investigate the spatial structure of a
protoplanetary disc whose dynamics is governed by magnetorotational turbulence.
We perform a series of local 3D chemo-radiative MHD simulations located at
different radii of a disc which is twice as massive as the standard minimum
mass solar nebula of~\cite{Hayashi1981}.  The ionisation state of the disc is
calculated by including collisional ionisation, stellar X-rays, cosmic rays and
the decay of radionuclides as ionisation sources, and by solving a simplified
chemical network which includes the effect of the absorption of free charges by
$\upmu$m-sized dust grains.  In the region where the ionisation is too low to
assure good coupling between matter and magnetic fields, a non-turbulent
central ``dead zone'' forms, which ranges approximately from a distance of 2 AU
to 4 AU from the central star.  The approach taken in the present work allows
for the first time to derive the global spatial structure of a protoplanetary
disc from a set of physically realistic numerical simulations.
\end{abstract}
\section{Introduction}
According to the present understanding, the process of planet formation takes
place in protoplanetary discs, which are accretion discs around young,
solar-type stars of the T~Tauri class.  Obtaining knowledge about the physical
conditions in protoplanetary discs is therefore of essential importance for
developing viable theories of how planets may form.

One of the most intriguing fact about protoplanetary discs is their rather short
lifetime of only about 10 million years \citep{Hartmann1998}.  The big question
is: How can the matter in the disc get rid of its angular momentum in such a
short time-scale?  The most likely explanation for this is hydromagnetic
turbulence initiated by the magnetorotational instability (MRI), a process which
was introduced into the context of accretion disc physics by
\cite{BalbusAndHawley1991}.  Numerical simulations show that magnetorotational
turbulence leads to fast outward angular momentum transport that can explain the
high accretion rates observed for protoplanetary discs \citep{KingEtAl07}.
However, the MRI will work properly only if there is good coupling between
matter and magnetic fields.  Since protoplanetary discs are cool objects, with
temperatures in the range of several hundred down to a few tens of Kelvin in
most parts of the disc, this is a critical issue.

Where in the disc the MRI is active, and where not, does depend on the value of
the ionisation level there.  While in the hot, inner regions (at a distance
$R<1\,\mathrm{AU}$ from the central star) collisional ionisation suffices to
provide good coupling, this is no longer true in the cooler regions further
away from the star.  There, the disc has to rely on other ionisation sources,
like the decay of radionuclides, cosmic rays and X-rays, which are emitted
from the corona of the central star.  In the planet-forming region, at
distances of several AU from the star, the density is so high that neither
the X-rays nor the cosmic rays are able to reach the midplane of the disc,
leading to a poorly ionised, non-turbulent central ``dead zone''
\citep{Gammie1996,Armitage2011}.  At these intermediate distances, only the
upper layers of the disc are expected to be turbulent and to still provide a
small amount of angular momentum transport.  In the outer regions, the surface
density is low enough for the X-rays to penetrate the whole disc column, so the
midplane becomes turbulent again.

While analytical calculations like in \cite{Gammie1996} or simplified 1+1D
models like the one developed in \cite{KretkeAndLin2010} can already provide
useful models for the structure of protoplanetary discs, the definite answers
on the questions of the size of the dead zone and the strength of the angular
momentum transport in the disc can only come from numerical simulations.  At
the present stage it seems very difficult to perform global simulations
including realistic physics due to the large computational cost and the
numerical complexity \citep[see][for examples of global protoplanetary disc
simulations]{FromangAndNelson2006,FromangAndNelson2009,DzyurkevichEtAl2010,FlockEtAl2011}.
On the other hand, the simpler local simulations, which model only a small part
of the disc, do already at the present time allow for the inclusion of
additional physics like radiation transport and disc chemistry
\citep{FlaigEtAl2010,HiroseAndTurner2011}.

In the present paper, we investigate the spatial structure of a protoplanetary
disc by performing a series of local 3D magnetohydrodynamical simulations
located at different radii, including both radiation transport and the effect
of a finite Ohmic resistivity.  We choose optimistic physical parameters in
order to obtain a small dead zone that fits inside the domain that is
simulated.  Using this method, we are able to obtain a comprehensive picture of
both the vertical and the radial structure of a magnetorotationally turbulent
protoplanetary disc. 

It should be noted that apart from Ohmic resistivity, ambipolar diffusion might
also reduce the saturation level of the MRI.  The strength of this effect
depends on the value of the neutral-ion collision frequency \citep[see, for
example,][]{BaiAndStone2011}.  In the fully turbulent regions of our model, the
ratio of collision frequency to orbital frequency is $\gtrsim$ 100, suggesting
that in these regions, the saturation level would not be strongly affected by
ambipolar diffusion.

The plan of our paper is as follows:  In Sec.~2 we describe our physical model
and the numerical setup.  Sec.~3 presents the results of the numerical
simulations and draws connections with astrophysical observations.  In Sec.~4,
we conclude.
\section{\label{ModelSetup}Model Setup}
Our basic setup is very similar to that of the radiative protoplanetary disc
simulations described in \cite{FlaigEtAl2010}.  The simulations take place in
the so-called stratified local shearing box, which is a rectangular box that
covers the full vertical height of the disc but has only a small radial and
azimuthal extent.  This allows the use of local Cartesian coordinates $(x,y,z)$,
where $x$ corresponds to the radial, $y$ to the azimuthal and $z$ to the
vertical direction, respectively \citep[for more information on the shearing box
setup, see][]{HawleyEtAl1995,StoneEtAl1996,StoneEtAl2010}.  At the vertical
boundaries, outflow boundary conditions are applied, that allow matter and
radiation to escape from the disc \citep[see][]{FlaigEtAl2010}.

The disc gas is described by the equations of magnetohydrodynamics, where we
include radiation transport in the one-temperature flux-limited diffusion
approximation \citep{FlaigEtAl2010} as well as the effect of a finite Ohmic
resistivity.  The physical equations are then given by
\begin{gather*}
%------------------------------------------------------------------------------%
	\frac{\partial \rho}{\partial t} + \nabla \cdot (\rho \bm v) = 0, \\
%------------------------------------------------------------------------------%
	\frac{\partial (\rho \bm v)}{\partial t} +
		\nabla \cdot \left[ \rho \bm v \bm v 
		+ p_\mathrm{tot} \mathsf{I}
		- \frac{\bm B \bm B}{4 \uppi} \right]	
		= \bm f_\mathrm{ext}, \\
%------------------------------------------------------------------------------%
	\frac{\partial \bm B}{\partial t} - \nabla 
	\times (\bm v \times \bm B - \eta \nabla \times \bm B) = 0, \\
%------------------------------------------------------------------------------%
	\frac{\partial e_\mathrm{tot}}{\partial t} + \nabla \cdot \left[
		\left(e_\mathrm{tot} + p_\mathrm{tot} \right) \bm v
		- \frac{(\bm v \cdot \bm B)\bm B}{\mu_0} 
		+ \frac{\eta}{\mu_0^2} \nabla
		\times \bm B \times \bm B \right],
	\\
		= \bm f_\mathrm{ext} \cdot \bm v - \nabla \cdot \bm F;
%------------------------------------------------------------------------------%
\end{gather*} 
with the total energy $e_\mathrm{tot} = p/(\gamma - 1) + \rho
v^2/2 + B^2/2 \mu_0$, the total pressure $p_\mathrm{tot} = p +
B^2/2 \mu_0$,  and 
\begin{equation}
	\bm f_\mathrm{ext}
=
	- 2 \rho \varOmega \hat{\bm z} \times \bm v
	+ 3 \rho \varOmega^2 x \, \hat{\bm x}
	- \rho \varOmega^2 z \, \hat{\bm z}
\end{equation}
denotes the source terms arising in the local shearing box
frame~\citep{FlaigEtAl2010}, with $\varOmega$ the local orbital frequency.  The
radiation flux is given by $\bm F = -(\lambda c / \kappa \rho) \nabla aT^4$.
The use of the flux-limiter $\lambda$ \cite[for which we use the form
suggested by][]{LevermorePomraning1981} makes the radiation transport method
applicable also to optically thin regions.

The above equations are solved using a conservative finite volume scheme.  The
scheme employs the HLLD Riemann solver of \cite{MiyoshiAndKusano2006}, which
yields a high effective resolution at moderate computational cost.  

The value of the resistivity $\eta$ is calculated by including various
ionisation source and by solving a simplified chemical network.  As in
\cite{HiroseAndTurner2011}, the values for the resistivity are read from a
precomputed table.  The value of the resistivity depends on the density, the
temperature and the local ionisation rate due to X-rays, cosmic rays and the
decay of radionuclides. We now describe the prescription according to which the
resistivity is calculated.
\subsection{Ionisation sources}
\subsubsection{Collisional ionisation}
In the hot, inner region inside 1\,AU, collisional ionisation is the dominant
ionisation source.  The ionisation level arising from this process is 
calculated using the Saha equation, which is given by
\begin{multline}
	x_\mathrm{e}
=	\frac{n_\mathrm{e}}{n_\mathrm{n}}
=	6.47 \cdot 10^{-13}
\left(\frac{T}{10^3}\right)^\frac{3}{4} \\
\times \left(\frac{2.4\cdot10^{15}}{n_\mathrm{n}}\right)^\frac{1}{2}
\frac{\exp(-25\,188/T)}{1.15\cdot10^{-11}},
\end{multline}
where $n_\mathrm{e}$ and $n_\mathrm{n}$ are the electron and neutral number
densities in cm$^{-3}$, respectively, $T$ is the temperature given in Kelvin and
we have assumed a potassium abundance of 10$^{-7}$ \citep[see, for
example][]{FromangEtAl2002}.
\subsubsection{Stellar X-rays}
In the region where collisional ionisation is low, stellar X-rays are the
dominant ionisation source.  For the ionisation rate due to X-rays, we use the
formula as given by \cite{TurnerAndSano2008}, 
\begin{multline}
	\zeta_\mathrm{XR}(R,z) = 2.6 \cdot 10^{-15} \, \mathrm{s^{-1}}
		\left( \frac{R}{1\,\mathrm{AU}} \right)^{-2}
		\left( \frac{L_\mathrm{XR}}{2 \cdot 10^{30}\,\mathrm{erg\,s^{-1}}} \right)
\\
		\times \left\{ \exp 
			\left( \frac{-\varSigma^+(R,z)}{8\,\mathrm{g\,cm^{-2}}} \right)
		+	\left( \frac{-\varSigma^-(R,z)}{8\,\mathrm{g\,cm^{-2}}} \right)
		\right\},
\end{multline}
where $\varSigma^{\pm}(R,z)$ are the column densities at radius $R$ above and
below a given point $z$,
\begin{equation}
	\varSigma^{\pm}(R,z)
=	\pm \int_z^{\pm\infty} \rho(R,z') \, \mathrm dz',
\end{equation}
and $L_\mathrm{XR}$ is the stellar X-ray luminosity which we take as
\begin{equation}
	L_\mathrm{XR}
=	2 \cdot 10^{31}.
\end{equation}	
Although this is a rather optimistic value, it is still inside the usual range
of $10^{29}-10^{32}\,\mathrm{erg\,s^{-1}}$ for the X-ray luminosities observed
for young stellar objects.
\subsubsection{Cosmic rays}
Another possibly important ionisation source are cosmic rays, i.e. highly
energetic particles from the interstellar medium that hit the disc.  If these
particles can reach the interior of the disc, their contribution is given by
\citep{2009ApJ...690...69U}
\begin{multline}
	\zeta_{\mathrm{CR}}(R,z) = \frac{\zeta_{\mathrm{CR},0}}{2} 
	\left\{  \phantom{
	\exp\left( -\frac{\varSigma^+(R,z)}{\chi_\text{CR}} \right) 
	\left. \left[ 1 + \left( \frac{\varSigma^+(R,z)}{\chi_\text{CR}} \right)^{\frac{3}{4}}
	\right]^{-\frac{4}{3}} \right.} \right. \\
	\exp\left( -\frac{\varSigma^+(R,z)}{\chi_\text{CR}} \right) 
	\left. \left[ 1 + \left( \frac{\varSigma^+(R,z)}{\chi_\text{CR}} \right)^{\frac{3}{4}}
	\right]^{-\frac{4}{3}} \right. \\
	+ \left. \exp\left( -\frac{\varSigma^-(R,z)}{\chi_\text{CR}} \right) 
	\left[ 1 + \left( \frac{\varSigma^-(R,z)}{\chi_\text{CR}}
	\right)^{\frac{3}{4}} \right]^{-\frac{4}{3}} \right\},
\label{eq:ionise:cr}
\end{multline}
where $\zeta_{\mathrm{CR},0}$ is set to the cosmic ray ionisation rate in the
interstellar medium, $\zeta_{\mathrm{CR},0} = 10^{-17} \, {\mathrm s^{-1}}$, and
$\chi_\text{CR} = 96.0 \, {\mathrm{g \, cm^{-2}}}$.
\subsubsection{Decay of radionuclides}
Finally, we also include the ionisation arising from the decay of
radionuclides.  Long-lived radionuclides provide a background ionisation level
of about $10^{-22} \, \mathrm{s^{-1}}$, where the dominant contribution is due
to $^{40}\text{K}$.  This rate can be significantly increased if one includes
the effect of short-lived radionuclides, which are no more present in the solar
system.  The most important contribution comes from the decay of
$^{26}\text{Al}$, which yields an ionisation rate of $\zeta_\mathrm{RA} = 9.2
\cdot 10^{-20} \, \mathrm{s^{-1}}$ based on the mean interstellar abundance and
eight times this value for the projected abundance of the young solar system,
which is based on the ratio of different aluminum isotopes found in CAIs in
meteorites \citep{2009ApJ...690...69U}.  Here we adopt an ionisation rate of
$\zeta_\mathrm{RA} = 7 \cdot 10^{-19} \, \mathrm{s^{-1}}$.
\subsection{Chemical network}
In the region where collisional ionisation is ineffective, the ionisation state
of the disc gas is determined by a balance between ionisation due to the
ionisation sources discussed above and recombination of free charges inside the
gas.
We use the extended Oppenheimer-Dalgarno network of \cite{2006A&A...445..205I},
which approximates the gas-phase chemistry by a generalized molecule species m
and metal species M,  extended by the introduction of spherical dust grains
which have a grain mass density of $\rho_\mathrm{gr} = 3$ $\mathrm{g\,cm^{-3}}$.
This extended network is called the ``ODD network'' in the following.

The charges of the newly introduced dust grains are tracked in the range from
$-2$ to $2$, leading to five new species $\mathrm{gr}^{2-}$ to
$\mathrm{gr}^{2+}$.  Furthermore, two mantle-species are introduced m(gr) and
M(gr), which represent molecules and metals frozen out on the grain surfaces,
leading to a total number of 12 species.

\begin{figure}
  \centering
  \includegraphics[scale=0.6]{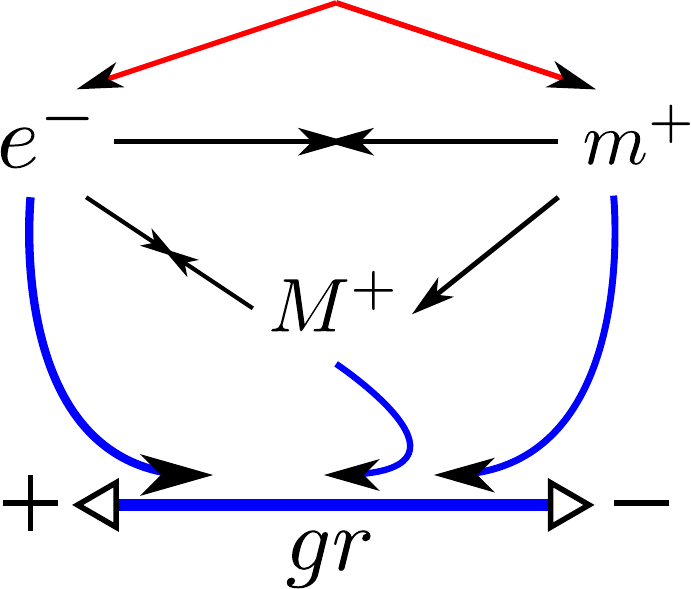}
  \caption{Schematic diagram of the charge-transfer reactions in the ODD
network.  The symbols m, M and gr denote the generalised molecule species (i.e.
mainly H$_2$), the generalised metal species (mainly K) and the dust grains,
respectively.  The red arrows denote the ionisation of molecules, the black
arrows denote the recombination and charge transfer reactions involving only
molecules and metals, while the blue arrows denote the absorbtion of charged
particles on the surface of dust grains.  \label{fig:odd:rscheme}}
\end{figure}

The reactions among the different species can be found in \citep[table 1, 3 and
4]{2006A&A...445..205I} and are  graphically illustrated in
Fig.~\ref{fig:odd:rscheme}.  (As we are mainly interested in the overall
ionisation fraction, only the reactions involving charged particles are shown).
The red ionisation reaction and the black charge transfer and recombination
reactions are the same as in the \cite{OppenheimerDalgarno1974} network, whereas
the reactions involving dust grains are drawn in blue.

The resistivity is calculated from the ionisation fraction according to the
usual formula
\begin{equation}
    \eta
=   234 \, x_\mathrm{e} / \sqrt{T}.
\end{equation}
The main effect of the dust grains is to provide new, indirect recombination
paths for electrons: The grains sweep up free electrons, charging up negatively
and simultaneously acquire positive charge through charge transfer reactions
with metal and molecule ions.  Grain-grain charge transfer reactions ensure that
oppositely charged grains neutralise themselves resulting in a relatively narrow
grain charge distribution.  In contrast to the metal ions, which provide a
charge reservoir that can only have a significant effect if enough metal is
present in the gas, dust grains behave as a kind of catalyst for recombination.
It is therefore possible that even tiny amounts of dust grains can change the
equilibrium electron fraction significantly.
\subsubsection{Equilibration timescale}
\begin{figure}
  \scalebox{0.7}{ 
	\includegraphics[width=0.65\textwidth]{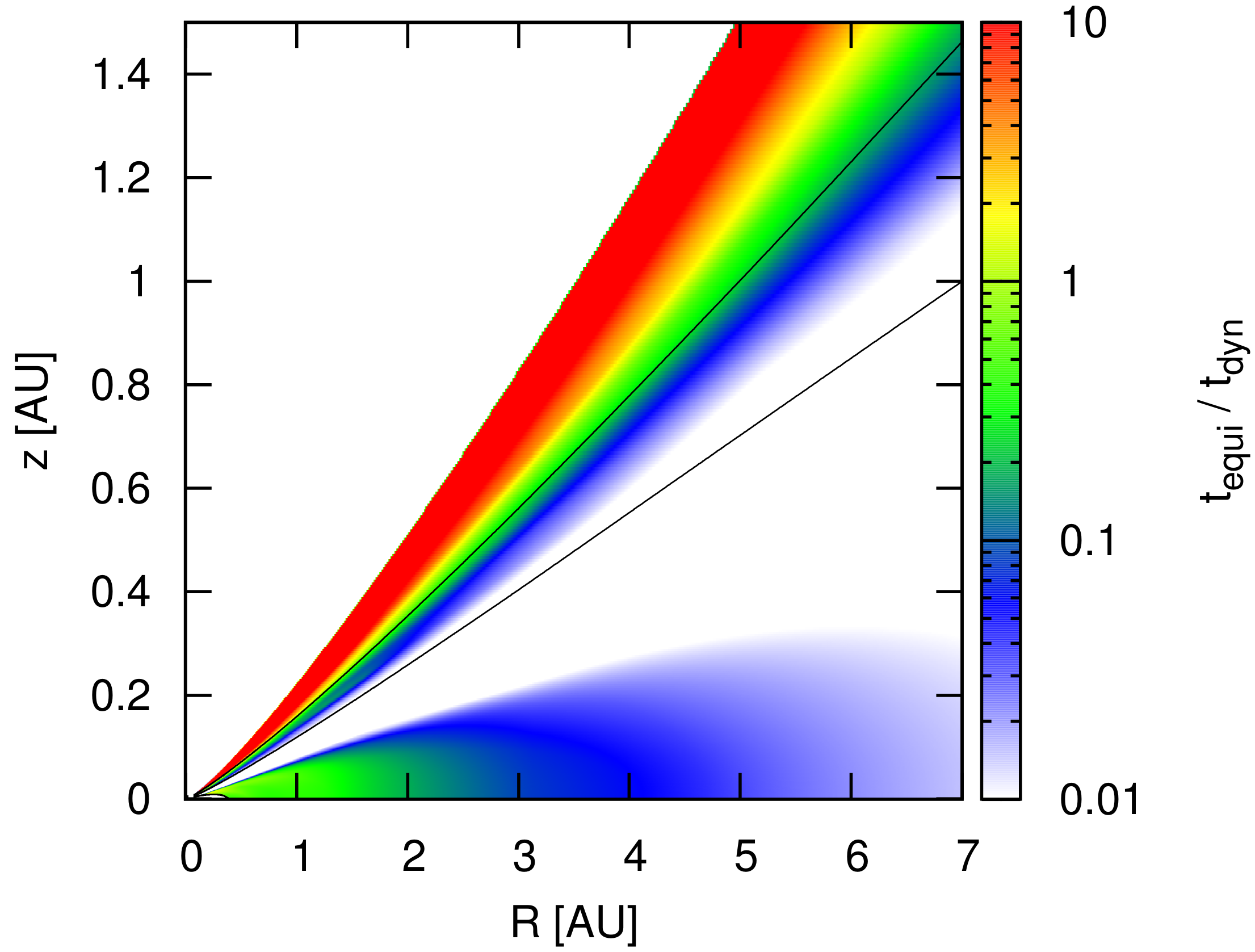}
  }
  \caption{Equilibration time of the ODD network over orbital time as a function
of the disc position.  The contour lines correspond to a magnetic Reynolds
number of $10^2$ and $10^4$, respectively.  For further comments, see the
text.\label{fig:odd:t_equi}}
\end{figure}

Fig.~\ref{fig:odd:t_equi} shows a plot of the equilibration timescale
$t_\mathrm{equi}$ over the orbital time $t_\mathrm{dyn} = 2 \uppi / \varOmega$
for the ODD network, with a grain size of 10 $\upmu$m, a dust-to-gas ratio of
$10^{-4}$, a metallicity of $10^{-10}$ and a disc temperature that varies with
the distance $R$ to the central star according to
\begin{equation}
T(R) = 280\, (R/\mathrm{AU})^{-1/2} \,\,\mathrm{K}.
\end{equation}
In addition, we also show contour lines for the magnetic Reynolds number defined
as
\begin{equation}
	\text{Re}_\text{m}
\defby
	\frac{H c_\mathrm{s}}{\eta}
=
	\frac{{c_\mathrm{s}}^2}{\eta\, \varOmega},
	\label{Re_m} 
\end{equation}
where $c_\mathrm{s}$ is the sound speed and $H=c_\mathrm{s}/\varOmega$ the
pressure scale height.  The critical value of $\text{Re}_\text{m}$ for which
the transition from the MRI-active to dead occurs, lies somewhere between
$10^2$ (for the case where the vertical magnetic flux through the disc is
non-zero) and $10^4$ (for a zero net flux configuration)
\citep[see][]{2006A&A...445..205I}.  As can be seen from
Fig.~\ref{fig:odd:t_equi}, the equilibration timescale mostly stays below the
orbital time in the region where the transition from MRI-dead zone to the
active zone occurs.  This means that turbulent mixing effects can be neglected
and the disc gas can be assumed to be in chemical equilibrium all the time.  We
note that when using a more complex chemical network, the equilibrium timescale
is expected to be even shorter, since there are then more recombination
channels available \citep{Bai2011}.  The chemical equilibrium value for the
electron concentration depends only on three parameters, the ionisation rate
$\zeta$, the temperature $T$ and the total number density of the gas $n$
(assuming a constant dust-to-gas mass ratio and metallicity). This fact allows
us to use a previously computed, three dimensional look-up table for the
equilibrium resistivity values in our MHD simulation instead of evolving the
chemical network during the simulation.
\subsection{Physical parameters}
We take the central star to have one solar mass.  Concerning the surface density
profile, we assume a disc mass which is two times the disc mass as given by the
minimum mass solar nebula model of~\cite{Hayashi1981}, i.e. the surface density
profile is given by the formula
\begin{equation} \label{2xHayashi}
   \varSigma_0
=  3400 \, \mathrm{g\,cm^{-2}}
   \left( \frac{R}{1\,\mathrm{AU}} \right)^{-3/2}.
\end{equation}
For this choice of surface density, the disc is to a large part optically thick
in the whole radial range spanned by our simulations (see Sec.~3).  This means
that the one-temperature approximation is well satisfied.

The disc gas is characterised by an adiabatic index of 1.4 and a mean molecular
weight of 2.35.  In accordance with astrophysical observations, we assume that
the dust particles in the disc have grown to a size of several microns, choosing
a monodisperse dust grain population with grain size $a_\mathrm{gr} = 10\,\upmu
\mathrm m$ for the chemical network.  We assume that the small dust grains are
partly depleted due to grain growth and assume a dust-to-gas ratio of $\rho_\mathrm{d}/\rho=10^{-4}$.
The metallicity is taken to be $M=10^{-10}$.   

As in \cite{FlaigEtAl2010}, we use the \cite{BellLin1994} opacity model.  Since
it is based on a dust grain population of smaller size than that used for the
chemical network, these two choices are not fully consistent.  This fact is,
however, not very serious, since the opacity does depend only weakly on the
grain size \citep{PollackEtAl1985} and the opacities in protoplanetary discs are
not well known anyway.  For this reasons, the use of an improved opacity model
is deferred to a future work.

For reference, all the basic physical parameters used in our model are summarised
in Table~\ref{tab:param}.
\begin{table}
\begin{minipage}{10cm}
	\begin{tabular}{lcc} 
\hline %------------------------------------------------------------------------
\hline %------------------------------------------------------------------------
	Parameter           & Symbol     & Value                        \\
\hline %------------------------------------------------------------------------
	Mass of central star        & $M_\star$   & $1\,M_\odot$                 \\
	Adiabatic index             & $\gamma$   & $1.4$                        \\
	Mean molecular weight       &$\mu$       & $2.35$                       \\
	X-Ray luminosity            & $L_\mathrm{XR}$   & $10^{31}\,\mathrm{erg\,s^{-1}}$                \\
	Cosmic ray ionisation rate  & $\zeta_\mathrm{CR}$  & $10^{-17}\,\mathrm{s^{-1}}$                 \\
	Ionisation due to radionuclides & $\zeta_\mathrm{RA}$  & $7 \cdot 10^{-19}\,\mathrm{s^{-1}}$\\
	Dust grain size & $a_\mathrm{gr}$  & $10\,\upmu \mathrm m$\\
	Dust-to-gas ratio  & $\rho_\mathrm{d}/\rho$  & $10^{-4}$ \\
	Material density of dust grains  & $\rho_\mathrm{gr}$  & 3 g\,cm${}^{-3}$ \\
	Metallicity  & $M$  & $10^{-10}$ \\
\hline %------------------------------------------------------------------------
	\end{tabular}
\end{minipage}
	\caption{\label{tab:param}Summary of basic physical parameters of the
physical model.  For further comments see the text.}
\end{table}
\section{Simulation Results}
We perform seven simulations located at different radii, starting from a
distance of 1\,AU from the central star up to a distance of 7\,AU.  The
simulations are initialised using the same set of initial conditions as
described in \cite{FlaigEtAl2010}, with an initial constant temperature $T_0$
(see Table~\ref{table-runs}), a hydrostatic (Gaussian) density distribution
with a surface density according to Eq.~\eqref{2xHayashi} and a zero net flux
magnetic field that has an initial plasma beta of $\beta=100$ at the midplane.
In terms of the initial pressure scale-height, the box size is one scale-height
in the radial direction and 6 scale-heights in the azimuthal direction.  The
vertical box size is 14 scale-heights for the simulations at $R=1$ AU and 10
scale-heights for the simulations at $R=5$ and $R=6$ AU, while for all other
simulations it is 8 scale-heights.  The corresponding box sizes as measured in
AU can be found in Table~\ref{table-runs}.  The simulations are seeded with
small random velocity perturbations of order $\sim 0.01 c_\mathrm{s}$, leading
to rapid development of the MRI, such that the disc reaches a fully turbulent
state already within the first ten orbits (see Fig.~\ref{fig:StressHist} later
in the paper).
\begin{table*}
	\begin{tabular}{cccccccccc} 
\hline %------------------------------------------------------------------------
\hline %------------------------------------------------------------------------
$R$ [AU]&\parbox{1.8cm}{Box size [AU]\\$L_x \times L_y \times L_z$}&
\parbox{1.8cm}{Resolution\\$n_x \times n_y \times n_z$}&Runtime [y]&
$T_0$ [K]&
$\frac{t_\mathrm{cool}}{t_\mathrm{orb}}$ & $H$ [AU] &$\langle \langle T \rangle \rangle$ [K]&
$\langle \langle \mathrm{Re}_\mathrm{m} \rangle \rangle$ &
$\langle \langle \alpha \rangle \rangle$\\
\hline %------------------------------------------------------------------------
1 & 0.07x0.41x0.97 & 32x64x384 & 100  & 1200 & 42.3  & 0.07  & 1537  & $1.02\cdot 10^{6}$ & 0.038 \\
2 & 0.11x0.67x1.35 & 32x64x256 & 283  & 400  & 29.8  & 0.09  & 249.7 & 441.5  & $9.3\cdot10^{-5}$ \\
3 & 0.15x0.88x1.17 & 32x64x256 & 519  & 200  & 24.0  & 0.11  & 139.0 & 515.9  & 0.001  \\
4 & 0.22x1.35x1.80 & 32x64x256 & 800  & 200  & 6.5   & 0.15  & 122.5 & 1091.8 & 0.004  \\
5 & 0.31x1.89x3.15 & 32x64x256 & 1118 & 200  & 2.3   & 0.28  & 161.8 & 24903.4& 0.013  \\
6 & 0.36x2.15x3.58 & 32x64x256 & 1469 & 150  & 6.7   & 0.34  & 163.1 & 17336.4& 0.039  \\
7 & 0.45x2.71x3.61 & 32x64x256 & 1851 & 150  & 3.3   & 0.24  & 54.4  & 1983.14& 0.011  \\
\hline %------------------------------------------------------------------------
		\end{tabular}
	\caption{\label{table-runs}Overview of the simulation runs that were
performed.  Listed are the distance to the central star (first column), the box
size (second column), the number of grid cells in each direction (third column),
the time in years each simulations has run (fourth column), the initial
temperature (fifth column), the initial cooling time in units of the orbital
period (sixth column), the final scale height (seventh column), the mean
temperature (eighth column), the mean magnetic Reynolds number (ninth column)
and the alpha parameter (tenth column).  The time-averages for the quantities
listed in columns eight to ten have been performed from 40 to 100 orbits.}
\end{table*}
Each simulation ran for 100 local orbits.  The corresponding runtimes in years
are also listed int Table~\ref{table-runs}.
\subsection{Time history}
\subsubsection{Thermal equilibrium}
In our setup, where we neglect the heating due to the irradiation from the
central star, the thermal structure of the disc is determined by a dynamical
balance between internal heating\footnote{ Note that the heating due to the
turbulent dissipation of kinetic and magnetic energy is included in this, since
by virtue of the conservative nature of the underlying numerical scheme, any
loss of kinetic and magnetic energy is automatically captured as gas internal
energy \citep[see][]{FlaigEtAl2010}.} and radiative cooling.  Estimating the
cooling time due to radiative diffusion from the relation
\begin{equation}
	t_\mathrm{cool}
=	H^2/D_\mathrm{rad},
\end{equation}
where the radiative diffusion coefficient $D_\mathrm{rad}$ is given by
\begin{equation}
	D_\mathrm{rad}
=	\frac{4acT^3}{3c_\mathrm{V}\kappa\rho^2}
\end{equation}
\citep[][]{FlaigEtAl2010}, we expect the simulations to reach thermodynamical
equilibrium after a few tens of orbits (see Table~\ref{table-runs}). The
resulting temperature should then be independent of the initial temperature.

\begin{figure} 
	\includegraphics[width=0.45\textwidth]{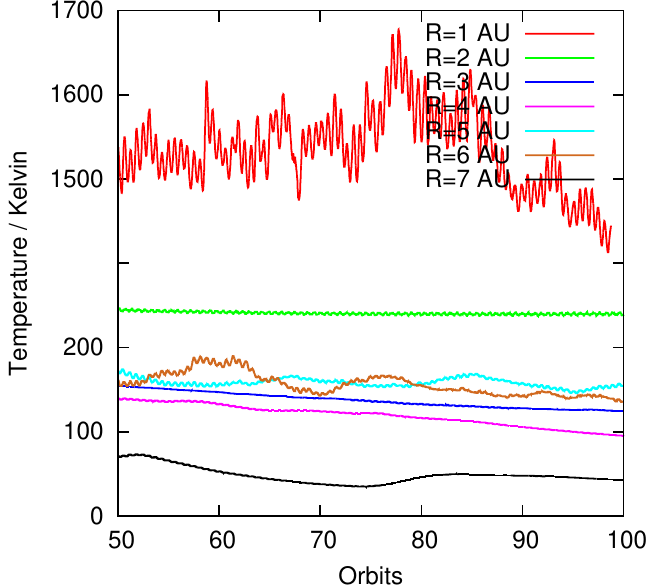}
	\begin{picture}(0,0)
		\put(-19,114){
			\begin{turn}{-45}
				\colorbox{white}{\begin{picture}(10,0)\put(-7,-5){\Huge=}\end{picture}}
			\end{turn}
		}
		\put(-196,116){
			\begin{turn}{-45}
				\colorbox{white}{\begin{picture}(10,0)\put(-7,-6){\Huge=}\end{picture}}
			\end{turn}
		}
	\end{picture}
	\caption{\label{fig:EthermHist}Density-weighted temperature as a function of
time.  Note that the $y$-axis has been broken in order to include the results
from the model at $R=1$ AU, which has a much higher temperature than the other
models.}
\end{figure}
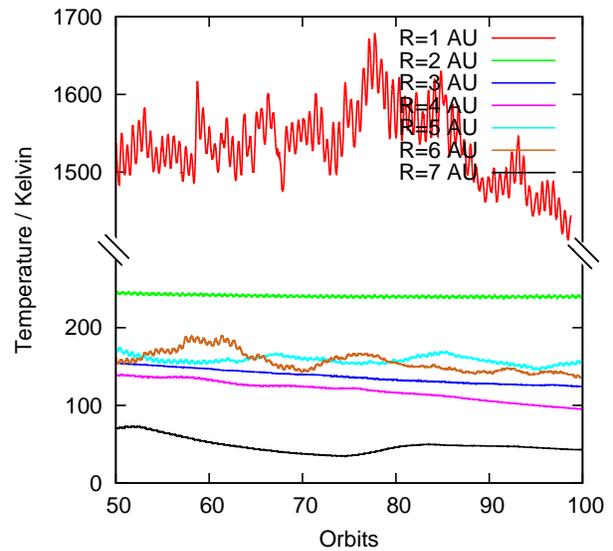

In Fig.~\ref{fig:EthermHist}, the temporal evolution of the density-weighted
spatially averaged temperature 
\begin{equation}
	\langle T \rangle_\rho
=
	\frac{\int \rho \, T \, \mathrm dV}{\int \rho \, \mathrm dV}
\end{equation}
is shown.  This plot suggest that the simulations do indeed reach a state of at
least approximate thermodynamical equilibrium, with the possible exception of
the simulations at $R=3$ and $R=4$ AU, which show a weak trend to cool during
the whole course of the simulation.

The time-averaged values of the temperature found in in the different
simulations are listed in Table~\ref{table-runs}.  Especially noticeable is the
sharp drop in temperature between the simulations at $R=1$ and $R=2$ AU, which
is due to the formation of a dead zone (see below).  
\subsubsection{Turbulent activity}
We measure the turbulent activity by calculating the $r$-$\phi$ (or $x$-$y$)
component of the stress tensor normalised to the gas pressure, the so-called
``alpha parameter'', according to the following prescription:
\begin{equation}
\langle \alpha \rangle
\equiv
\frac{\langle T_{xy} \rangle_\rho}{\langle p \rangle_\rho},
\end{equation}
where $T_{xy} = T^\mathrm{Reyn}_{xy} + T^\mathrm{Maxw}_{xy}$ and
$T^\mathrm{Reyn}_{xy}$ and $T^\mathrm{Maxw}_{xy}$ are the Reynolds and
Maxwell stresses, which are defined as
\begin{equation}
T^\mathrm{Reyn}_{xy}
=
\rho v_x \, \delta v_y
\quad
\mbox{and}
\quad
T^\mathrm{Maxw}_{xy}
=
-B_x B_y / \mu_0
\end{equation}
where $\delta v_y$ is the azimuthal component of the gas velocity with the
velocity of the background shear flow subtracted.  The quantity $T_{xy}$ is
related to the amount of outward angular momentum transport taking place at a
certain location \citep[see, e.g.][]{BalbusRev2003}.

\begin{figure} 
	\includegraphics[width=0.45\textwidth]{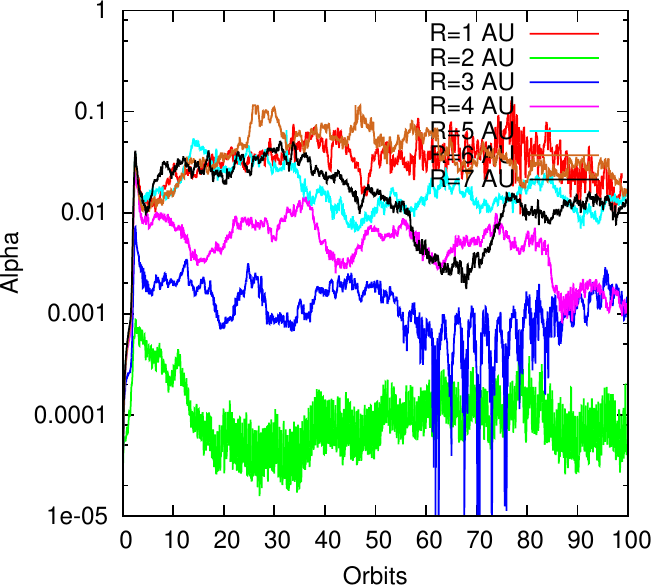}
	\caption{\label{fig:StressHist}Alpha parameter (turbulent stress normalised
to gas pressure) as a function of time.  For further comments, see the text.}
\end{figure}

The time evolution of the alpha parameter is plotted in
Fig.~\ref{fig:StressHist}, the time-averaged values of alpha can be found in
Table~\ref{table-runs}.
Within the first ten orbits, all simulations reach a state of saturated
turbulence and remain turbulent until the end of the simulation.

The value of alpha drops sharply when going from $R=1$ AU to $R=2$ AU, where
the ionisation level is very low, since the gas is too cool for collisional
ionisation and too thick for either the the X-rays or the cosmic rays too reach
the midplane (see Fig.~\ref{Fig:ExtIonSrc} later in the paper).  When going
further outwards, the ionisation level increases, since there a significant
fraction of the cosmic rays is able to penetrate the disc, leading to an
increase in the turbulent activity.  The simulation at $R=6$ AU is again fully
turbulent, with a mean alpha that is the same as for the simulation at $R=1$
AU.  The stresses found in the poorly ionised models at $R=2-4$ AU are one to
three orders of magnitude smaller than found for the well ionised case.  

Turbulence mixes gas and dust and counteracts the force of gravity which causes
the dust particles to settle towards the midplane.  Based on the levels of
turbulence found in our simulations, we can estimate up to which height gas and
dust can be considered to be well mixed.  We do this by equating the settling
time-scale $t_\mathrm{s} = 1/\varOmega^2 t_\mathrm{d}$ (where $t_\mathrm{d} =
\rho_\mathrm{gr} a_\mathrm{gr} / \rho c_\mathrm{s}$ is the drag time) with the
turbulent mixing time-scale $t_\mathrm{m} = H^2/D_\mathrm{dust}$. The dust
diffusion coefficient $D_\mathrm{dust}$ is related to the alpha parameter via
the Schmidt number $S_\mathrm{c}$ according to $S_\mathrm{c} = \alpha
c_\mathrm{s} H / D_\mathrm{dust}$ \citep{FromangAndPapaloizou2006}.  For a
density profile in hydrostatic equilibrium, $\rho(z) = \rho_0 \exp(-z^2/2
H^2)$, the height $z_0$ up to which gas and dust are well mixed then follows
from $t_\mathrm{s} = t_\mathrm{m}$ to be
\begin{equation} \label{mixing-height}
	\frac{z_0}{H}
=	\sqrt{-2 \ln \frac{\alpha}{S_\mathrm{c}} \varOmega t_\mathrm{d}(z=0)}.
\end{equation}
Numerical simulations indicate a Schmidt number of order unity
\citep{JohansenAndKlahr2005,TurnerEtAl2006,FromangAndPapaloizou2006}.  Setting
$S_\mathrm{c} = 1$, and using the initial values for the midplane density and
sound speed, as well as the alpha values from Table~\ref{table-runs},
Eq.~\eqref{mixing-height} predicts $z_0 = 2.3 H$ for the run at $R=2$ while for
all other runs $z_0 > 3 H$.  Since the MRI operates mainly within the first
three scale-heights (cf. Fig.~\ref{fig:AlphaVert} later in the paper), this
means that gas and dust can be considered to be well mixed in the MRI active
regions, so the approximation of a constant dust-to-gas ratio is indeed
justified.
\subsection{Spatial structure}
\subsubsection{Ionisation level \& stress profiles}
\begin{figure} 
	\includegraphics[width=0.45\textwidth]{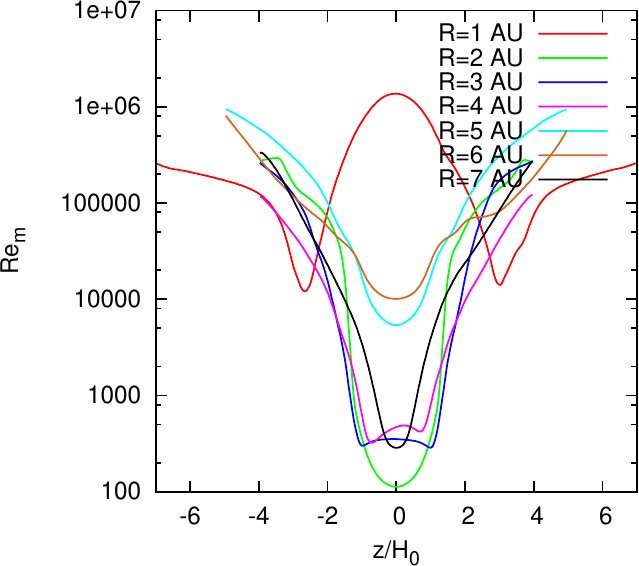}
	\caption{\label{fig:ReMVert}Vertical profiles of the mean magnetic Reynolds
number, averaged from 60 to 100 orbits.  The distance from the midplane is given
in units of the initial pressure scale-height $H_0$.  For further comments, see
the text.}
\end{figure}

Concerning the spatial structure of the disc, we first consider the question of
how the ionisation level and the turbulent acitvity vary as a function of
position.  In Fig.~\ref{fig:ReMVert}, we plot the vertical profile of the
magnetic Reynolds number, as defined in Eq.~\eqref{Re_m}.

Judging from the condition that regions with $\mathrm{Re_m} > 10^4$ can be
considered sufficiently ionised for the MRI not to be reduced, we can say that
the simulations at $R=1$ and $R=6$ AU can be considered as ideal.  All other
simulations contain at least a small zone of insufficient ionisation near the
midplane. 

\begin{figure} 
	\includegraphics[width=0.45\textwidth]{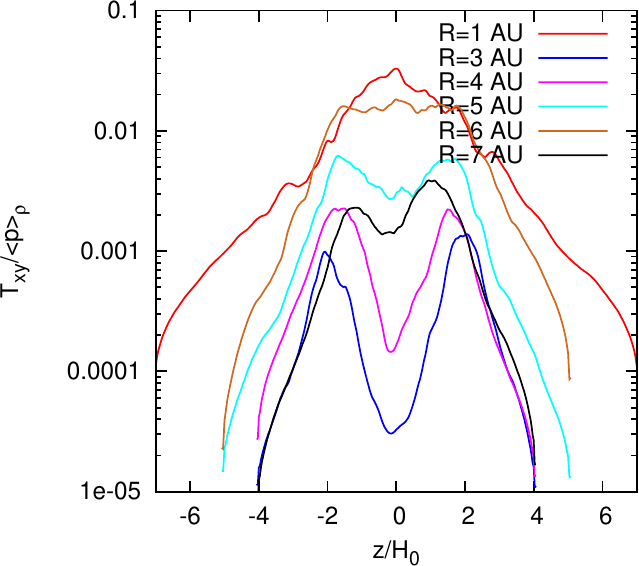}
	\caption{\label{fig:AlphaVert}Vertical profiles of the total stress
normalised to the mean gas pressure $\langle p \rangle_\mathrm{\rho}$, showing
the formation of non-turbulent zones near the midplane due to insufficient
ionisation.  Averages have been performed from 60 to 100 orbits.}
\end{figure}

Consequently, the vertical stress profiles, which are shown in
Fig.~\ref{fig:AlphaVert},
are roughly similar for the ``ideal'' simulations at $R=1$ and $R=6$ AU.  The
simulations at $R=5$ and $R=7$ AU exhibit only a small dip in the stress near
the midplane, while in the other simulations, the stresses drop noticeably
there, forming an extended non-turbulent dead zone ranging from a distance of
$R=2-4$ AU from the central star.

As in the paper of \cite{SimonAndHawley2011}, who also used the HLLD Riemann
solver, we do not observe a double peak in the stress profiles of the well
ionised simulations at $R=1$ and $R=6$ AU.  This is contrary to previous
previous stratified simulations like that of \cite{HiroseEtAl2009} or
\cite{FlaigEtAl2010}.  Since the only major difference in the numerical code as
compared to the calculations presented in \cite{FlaigEtAl2010} is the change of
Riemann solver, it seems that the absence of the double peak profile is indeed
due to the use of the HLLD solver.
\subsubsection{Structure of the magnetic field}
\begin{figure*} 
	\includegraphics[width=0.3\textwidth]{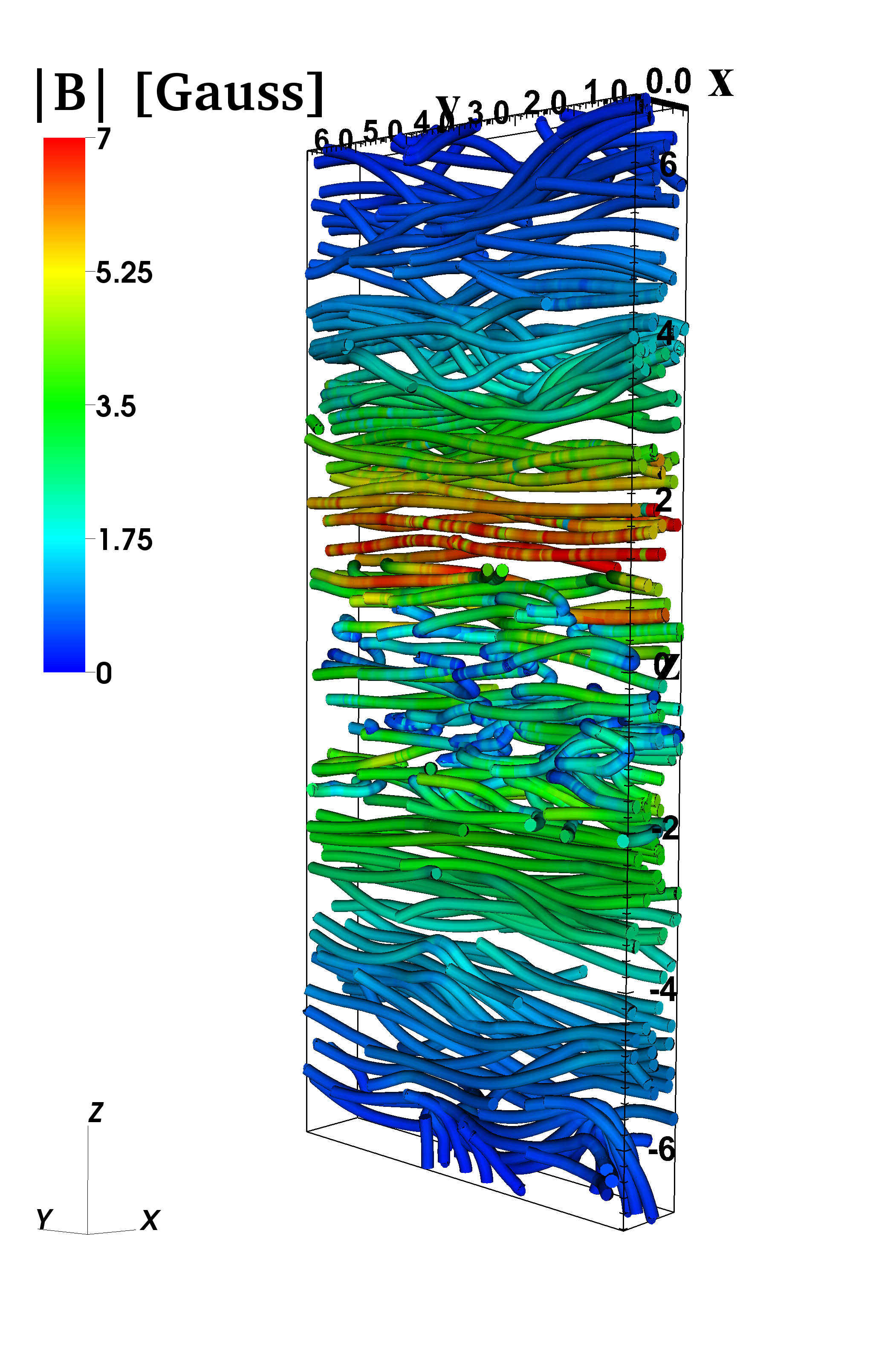}
	\includegraphics[width=0.3\textwidth]{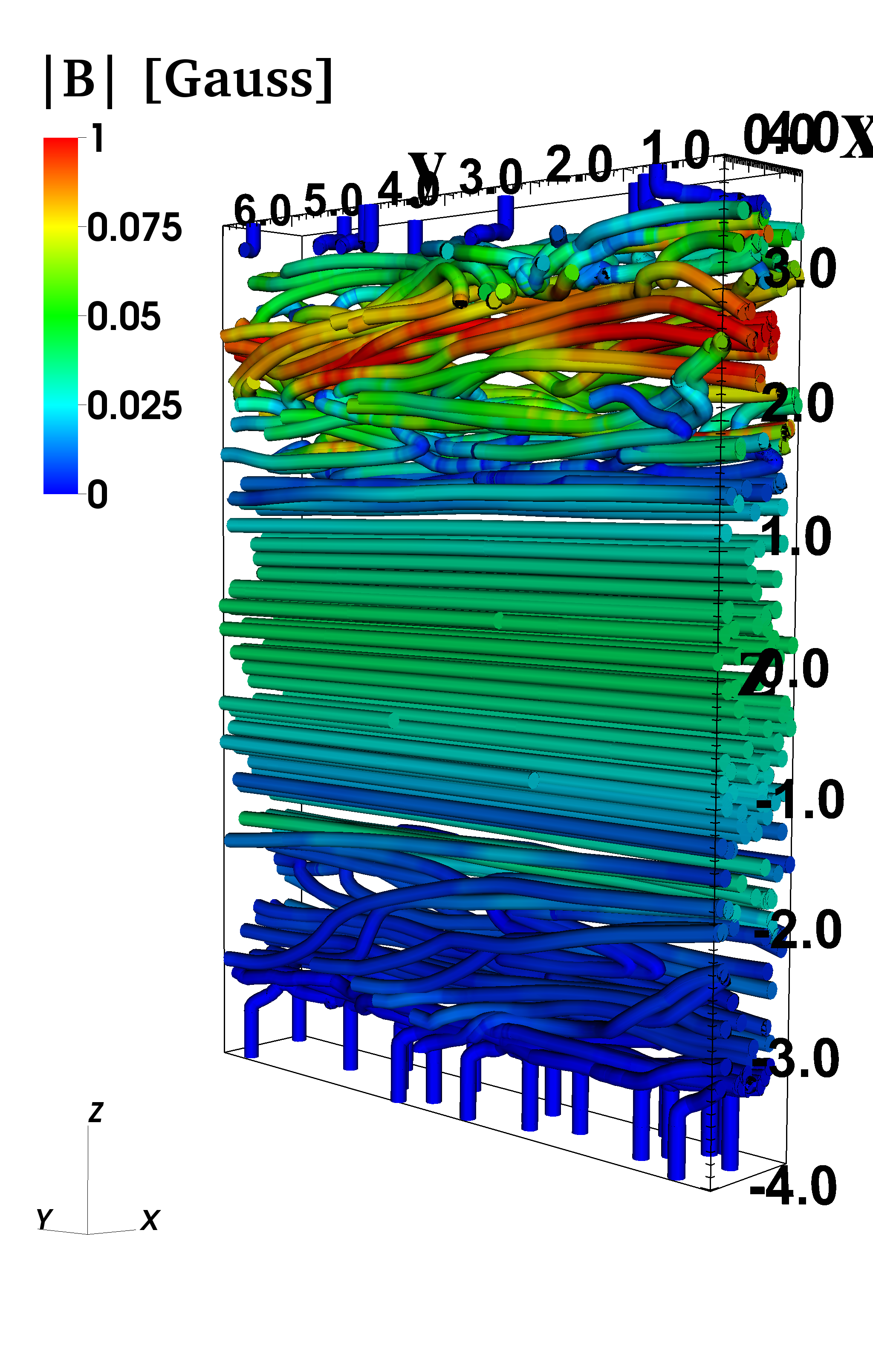}
	\includegraphics[width=0.3\textwidth]{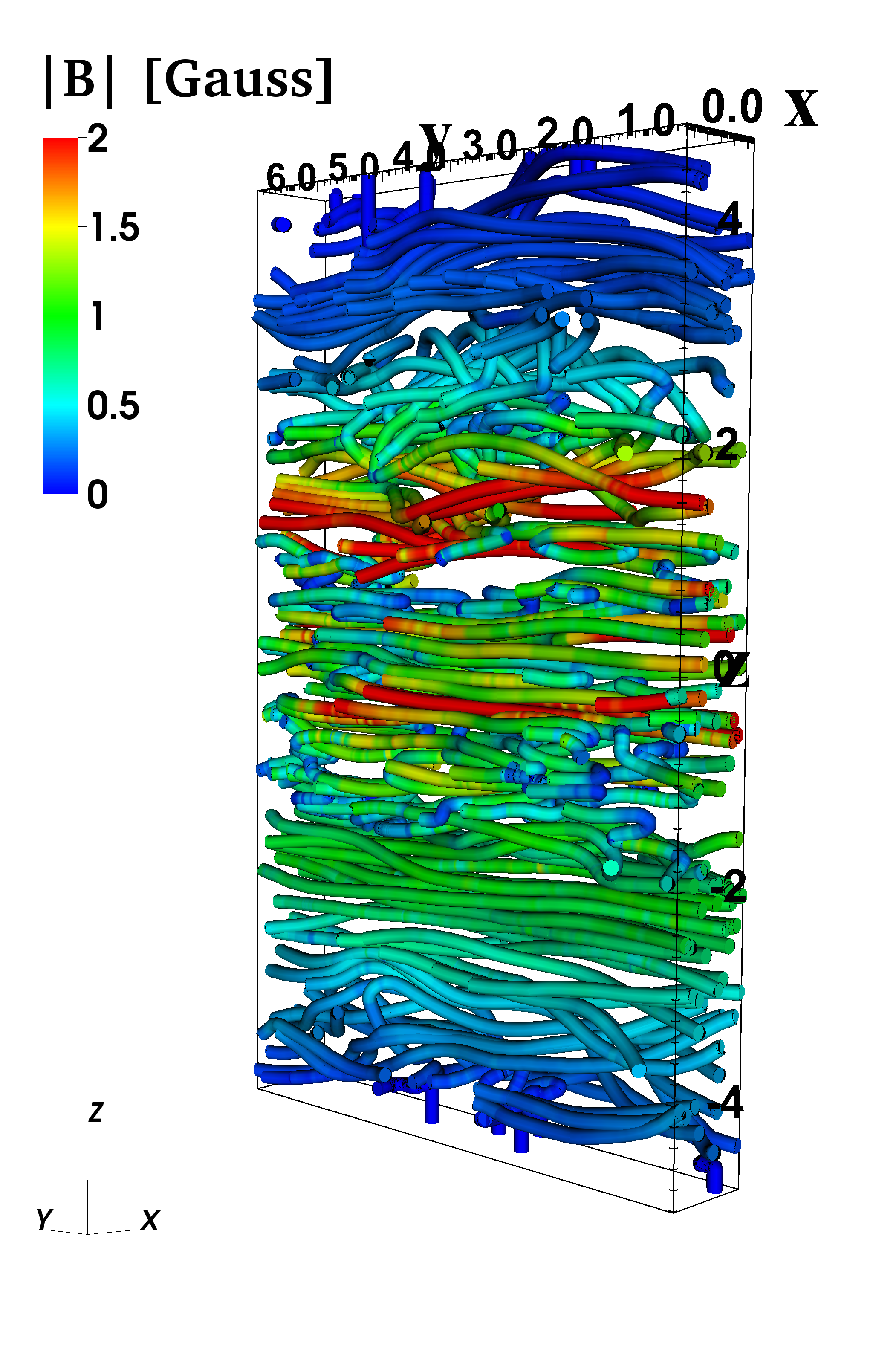}
	\caption{\label{fig:BFields}Snapshots of the magnetic field structure for
the runs at $R=1$ AU (left plot, taken at $t=50$ orbits), $R=2$ AU (middle plot, taken at
$t=90$ orbits), and $R=5$ AU (right plot, taken at $t=80$ orbits).  The
distances are given in terms of the initial pressure scale-height.}
\end{figure*}

Concerning the structure of the magnetic field, we plot snapshots of the
magnetic field structure for three different runs in Fig.~\ref{fig:BFields}.
The left plot shows the magnetic field structure in the fully ideal run at $R=1$
AU.  Within a distance of approximately two to three pressure scale-heights from
the midplane, the MRI is fully operational and the magnetic field is highly
tangled.  Further outwards, the disc becomes magnetically dominated, which leads
to a quenching of the MRI, with the magnetic field being predominantly
azimuthal.  Even further away from the midplane, above four pressure-scale
heights, the magnetic field becomes distorted again.  The magnetic field
structure that we find for the ideal model is very similar to the field
structure found by \cite{HiroseEtAl2006} \citep[see also][]{SimonAndHawley2011}.

In the non-ideal runs, the magnetic field structure looks quite different, as
one can see from the middle plot of Fig.~\ref{fig:BFields}, which shows the
magnetic field structure in the very poorly ionised run at $R=2$ AU.  Here the
magnetic field is laminar in the region around the midplane, with only the
upper layers retaining some level of turbulent activity.  This plot can be
compared with \cite{HiroseAndTurner2011}.

Finally, in the right plot of Fig.\ref{fig:BFields}, we show a snapshot of the
magnetic field for the run at $R=5$ AU.  The magnetic field structure is quite
similar to that of the run at $R=1$ AU, with the difference that the magnetic
field near the midplane is more laminar.
\subsubsection{Thermal structure}
\begin{figure} 
	\includegraphics[width=0.45\textwidth]{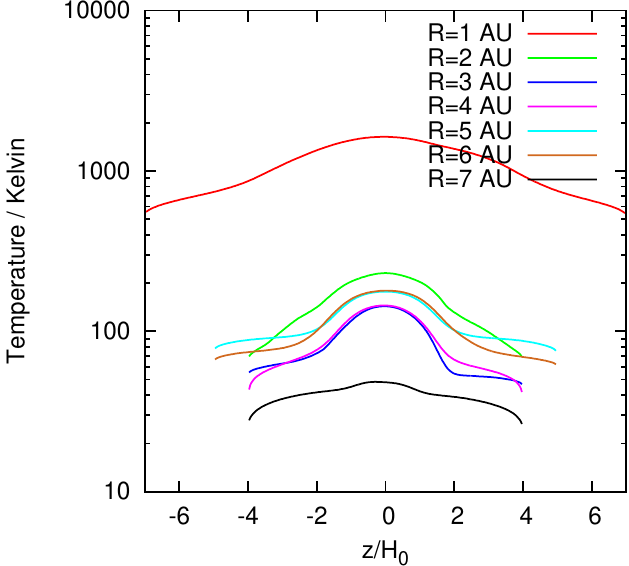}
	\caption{\label{fig:TempVert}Mean vertical temperature profiles, averaged
from 60 to 100 orbits.}
\end{figure}

Next, we turn to the question of how the thermal structure of the disc looks
like.  We show the temperature profiles in Fig.~\ref{fig:TempVert}. The
temperature decreases monotonically with increasing distance from the central
star except at the outer edge of the dead zone, where it increases when going
from $R=4$ AU to $R=5$ AU, due to the increased turbulent heating.  

\begin{figure} 
	\includegraphics[width=0.45\textwidth]{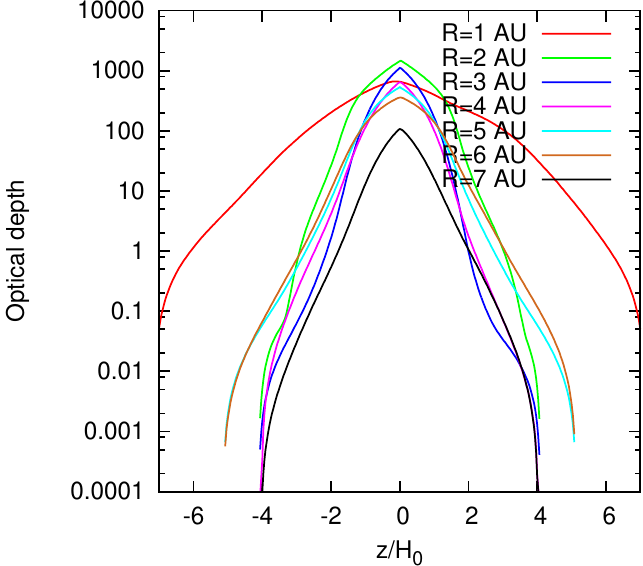}
	\caption{\label{fig:OpticalDepthVert}Mean vertical profiles of the optical
depth, averaged from 60 to 100 orbits.}
\end{figure}

Vertical profiles of the optical depth are shown in
Fig.~\ref{fig:OpticalDepthVert}.
In all the simulations, the midplane is optically thick, which justifies the use
of the one-temperature approximation.  Also, the photosphere is located well
inside the computational domain for all the simulations.

It is noticeable that the simulations at $R=2$ and $R=3$ AU have a larger
central optical depth than the simulation at $R=1$ AU.  Also, the optical depth
does not change very much for the five simulations in the range from 2 to 6 AU.
Both observations can be understood from the fact that these simulations lie in
a temperature range where the opacity increases strongly with decreasing
temperature (scaling like $\kappa \propto T^{-7}$ in this region), which
balances the effect of the change in surface density on the optical depth.
\subsubsection{Global disc structure}
By interpolating between the results obtained from the local simulations at
different radii, we can calculate time-averaged maps in the $R$-$z$ plane for
the quantities of interest, leading to an axisymmetric model of a protoplanetary
disc in steady state, which is derived from physically realistic
three-dimensional numerical simulations.

\begin{figure*} 
	\includegraphics[width=\textwidth]{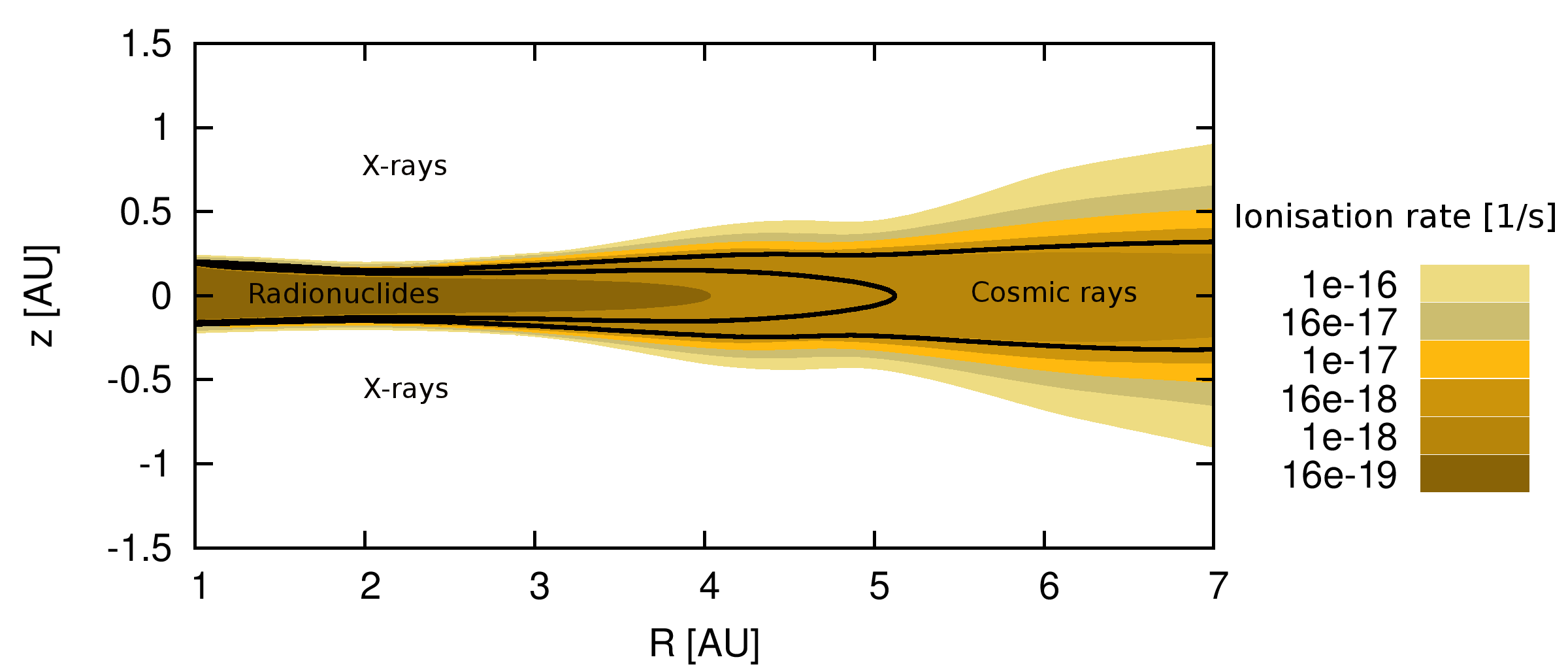}
	\caption{\label{Fig:ExtIonSrc}Ionisation rate due to the external
ionisation sources.  The black line corresponds to the boundary between regions
where different ionisation sources dominate.}
\end{figure*}

Fig.~\ref{Fig:ExtIonSrc} shows the total ionisation rate due to the stellar
X-rays, cosmic rays and the decay of radionuclides.  Within the first 3-4 AU,
the radionuclides dominate the ionisation rate near the midplane, while outside
of 4 AU, the cosmic rays dominate.  For the whole radial range considered, the
X-rays become dominant only in the upper layers, at distances $\gtrsim$ 0.2-0.3 AU away
from the midplane.

\begin{figure*}
	\includegraphics[width=\textwidth]{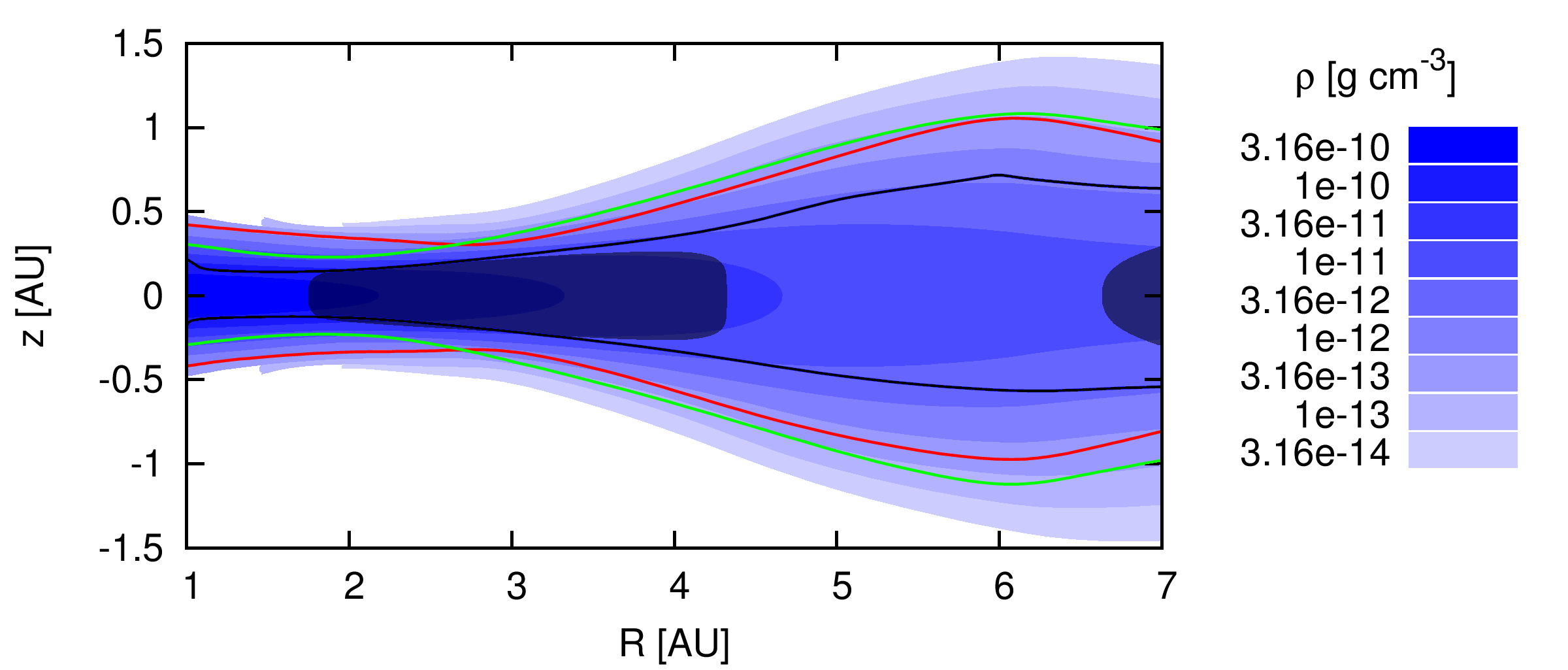}
	\caption{\label{fig:SpatialStructure}{\it Global spatial structure.} Shown
is a contour plot of the density.  The shaded area corresponds to the region
where the magnetic Reynolds number is smaller than 1000.  The red line marks the
location of the photosphere, the black line corresponds to the the location
where magnetic pressure and gas pressure are equal and the green line
corresponds to the boundary where the flow starts to become supersonic.}
\end{figure*}

Fig.~\ref{fig:SpatialStructure} is intended to give an impression of how the
disc looks on average in the quasi steady state obtained in the local
simulations presented in this paper.  Shown is the density, the location of the
photosphere and the magnetosphere as well as the location where the flow becomes
supersonic.  The average extent of the dead zone is also shown, based on the
criterion that the the magnetic Reynolds number be smaller than 1000.  The
appearance of a second dead zone at $R=7$ AU is likely an artefact of our
particular setup, where we neglect the heating due to the stellar irradiation,
leading to an unrealistically low temperature of only $\sim 50$ K and poor
ionisation.

We note that as in the model of \cite{FlaigEtAl2010}, which was located at
$R=1$ AU with a surface density about six times that of the minimum mass solar
nebula, the photosphere lies always inside the magnetically dominated region,
although the model presented here is much less massive, and we perform
simulations at larger radii.  As in the \cite{FlaigEtAl2010} model, inside the
first 2-3 AU, the turbulence is supersonic at the location of the photosphere,
but becomes slightly subsonic there at larger distances from the central star.
The conclusion drawn in \cite{FlaigEtAl2010} that the MRI can be considered as
a possible source for the turbulent line broadening observed in protoplanetary
discs, can thus in principle be carried over to models containing dead zones.
We note that a recent study \citep{SimonEtAl2011}, also finds supersonic
velocities above three scale-heights, both for ideal models and models
containing a dead zone.

\begin{figure*} 
	\includegraphics[width=\textwidth]{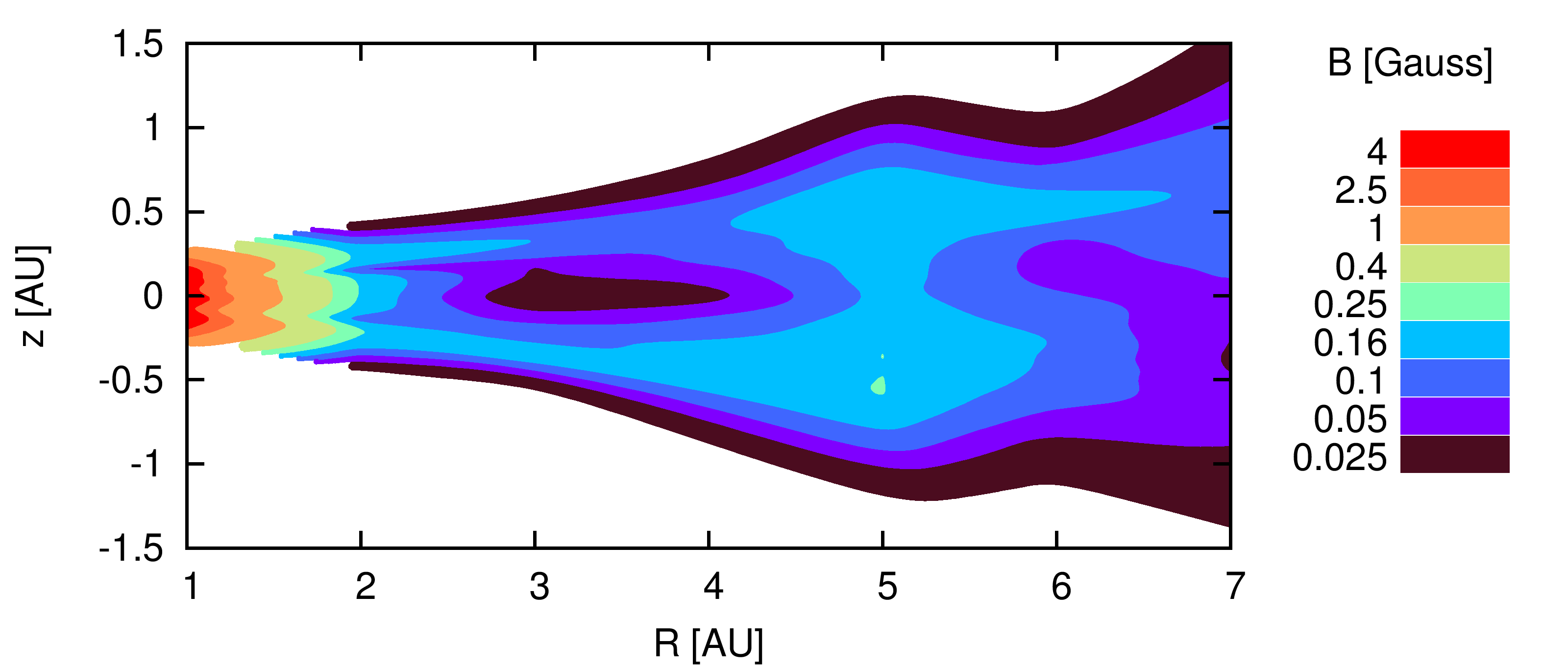}
	\caption{\label{fig:Babs2D}{\it Mean magnetic field strength.}}
\end{figure*}

Fig.~\ref{fig:Babs2D} shows the interpolated mean magnetic field strengths
found in the simulations presented in this paper.
The field strengths have of course a minimum at the location of the dead zone.
Outside of 2 AU, the mean field strengths are almost everywhere below $\approx
0.2$ G.  The field strengths obtained in the present model are probably too low
to explain the remnant magnetic fields found in chondritic meteorites, which
suggests that a more massive (probably fully active) disc is needed to explain
these observations \citep{KingAndPringle2010}.
\section{Summary \& Conclusion}
We have calculated the spatial structure of a magnetorotationally turbulent
protoplanetary disc from a series of physically realistic, local
three-dimensional numerical simulations.  Inside a distance of 1 AU from the
central star, the disc gas is sufficiently ionised by collisional ionisation, so
the disc midplane is fully turbulent.  Between 1 and 2 AU, the temperature drops
below the threshold of 900 K which is needed for collisional ionisation to be
effective, leading to a transition from the fully turbulent state to a state
that contains a central non-turbulent dead-zone.  When going further outwards,
the disc gas becomes thinner, which allows a larger fraction of the stellar
X-rays to reach the region around the midplane, leading to a gradual increase in
the ionisation level and also the turbulent activity.  At $R=6$ AU, the midplane
is again fully turbulent.

The disc gas is heated internally (via turbulent and Ohmic dissipation) with
the heating due to the stellar irradiation being neglected.  The simulations
reach (approximate) thermodynamical equilibrium, except possibly for the
simulation located at $R=3$ and $R=4$ AU, which continue to cool slowly for the
whole course of the simulation.  For these two particular runs, the temperature
may therefore dependent on the initial conditions, while for all the other
simulations, the temperature comes out self-consistently as a result of the
balance between internal heating and radiative cooling.  Since in most of our
simulations the disc gas is for a large part optically thick, we do not expect
the stellar irradiation to change the temperature in the interior of the disc
drastically, except for the simulation at $R=7$, where the temperature is
indeed unrealistically low \cite[when compared, for example, with the passive
disc model of][]{ChiangAndGoldreich1997}.

It should be noted that the use of the local approximation is a major
restriction of the present work, since local simulations cannot capture the
full dynamics of a global simulation \citep[see, for
example,][]{SorathiaEtAl2011}.  However, apart from the fact that in a global
setup the surface density would change over time, the spatial structure in a
global model using the same physics will likely not be dramatically different
from the structure coming out in the local framework.  Therefore, the approach
taken in the present work is a useful first step in deriving the structure of
protoplanetary discs from physically realistic numerical simulations.  

To summarise, the simulations presented in this work are able to capture all
the basic physics that is thought to be important for protoplanetary disc
dynamics, and do not suffer from the lack of self-consistency of previous
isothermal models.  For the first time, it is therefore possible to obtain a
picture of the global structure of a magnetorotationally turbulent
protoplanetary disc from first-principles numerical simulations.
\section*{Acknowledgements}
This research has been supported in part by the Deutsche Forschungsgemeinschaft
DFG through grant DFG Forschergruppe 759 ``The Formation of Planets: The
Critical First Growth Phase''.  Computational resources were provided by the
High Performance Computing Cluster of the University of T\"ubingen.  We thank
Neal Turner, Mario Flock, Natalia Dzyurkevich and Martin Ilgner for helpful
discussions.  We also want to thank the anonymous referee for very useful
suggestions.
\bibliography{FlaigEtAl2011}

\begin{thebibliography}{}

\bibitem[\protect\citeauthoryear{{Armitage}}{{Armitage}}{2011}]{Armitage2011}
{Armitage} P.~J.,  2011, \araa, 49, 195

\bibitem[\protect\citeauthoryear{{Bai}}{{Bai}}{2011}]{Bai2011}
{Bai} X.-N.,  2011, ArXiv e-prints

\bibitem[\protect\citeauthoryear{{Bai} \& {Stone}}{{Bai} \&
  {Stone}}{2011}]{BaiAndStone2011}
{Bai} X.-N.,  {Stone} J.~M.,  2011, \apj, 736, 144

\bibitem[\protect\citeauthoryear{Balbus \& Hawley}{Balbus \&
  Hawley}{1991}]{BalbusAndHawley1991}
Balbus S.,  Hawley J.,  1991, \apj, 376, 214

\bibitem[\protect\citeauthoryear{{Balbus}}{{Balbus}}{2003}]{BalbusRev2003}
{Balbus} S.~A.,  2003, \araa, 41, 555

\bibitem[\protect\citeauthoryear{{Bell} \& {Lin}}{{Bell} \&
  {Lin}}{1994}]{BellLin1994}
{Bell} K.~R.,  {Lin} D.~N.~C.,  1994, \apj, 427, 987

\bibitem[\protect\citeauthoryear{{Brandenburg}, {Nordlund}, {Stein} \&
  {Torkelsson}}{{Brandenburg} et~al.}{1995}]{BrandenburgEtAl1995}
{Brandenburg} A.,  {Nordlund} A.,  {Stein} R.~F.,    {Torkelsson} U.,  1995,
  \apj, 446, 741

\bibitem[\protect\citeauthoryear{{Chiang} \& {Goldreich}}{{Chiang} \&
  {Goldreich}}{1997}]{ChiangAndGoldreich1997}
{Chiang} E.~I.,  {Goldreich} P.,  1997, \apj, 490, 368

\bibitem[\protect\citeauthoryear{{Dzyurkevich}, {Flock}, {Turner}, {Klahr} \&
  {Henning}}{{Dzyurkevich} et~al.}{2010}]{DzyurkevichEtAl2010}
{Dzyurkevich} N.,  {Flock} M.,  {Turner} N.~J.,  {Klahr} H.,    {Henning} T.,
  2010, \aap, 515, A70+

\bibitem[\protect\citeauthoryear{{Flaig}, {Kley} \& {Kissmann}}{{Flaig}
  et~al.}{2010}]{FlaigEtAl2010}
{Flaig} M.,  {Kley} W.,    {Kissmann} R.,  2010, \mnras, 409, 1297

\bibitem[\protect\citeauthoryear{{Flock}, {Dzyurkevich}, {Klahr}, {Turner} \&
  {Henning}}{{Flock} et~al.}{2011}]{FlockEtAl2011}
{Flock} M.,  {Dzyurkevich} N.,  {Klahr} H.,  {Turner} N.~J.,    {Henning} T.,
  2011, \apj, 735, 122

\bibitem[\protect\citeauthoryear{{Fromang} \& {Nelson}}{{Fromang} \&
  {Nelson}}{2006}]{FromangAndNelson2006}
{Fromang} S.,  {Nelson} R.~P.,  2006, \aap, 457, 343

\bibitem[\protect\citeauthoryear{{Fromang} \& {Nelson}}{{Fromang} \&
  {Nelson}}{2009}]{FromangAndNelson2009}
{Fromang} S.,  {Nelson} R.~P.,  2009, \aap, 496, 597

\bibitem[\protect\citeauthoryear{{Fromang} \& {Papaloizou}}{{Fromang} \&
  {Papaloizou}}{2006}]{FromangAndPapaloizou2006}
{Fromang} S.,  {Papaloizou} J.,  2006, \aap, 452, 751

\bibitem[\protect\citeauthoryear{{Fromang}, {Terquem} \& {Balbus}}{{Fromang}
  et~al.}{2002}]{FromangEtAl2002}
{Fromang} S.,  {Terquem} C.,    {Balbus} S.~A.,  2002, \mnras, 329, 18

\bibitem[\protect\citeauthoryear{{Gammie}}{{Gammie}}{1996}]{Gammie1996}
{Gammie} C.~F.,  1996, \apj, 457, 355

\bibitem[\protect\citeauthoryear{{Hartmann}, {Calvet}, {Gullbring} \&
  {D'Alessio}}{{Hartmann} et~al.}{1998}]{Hartmann1998}
{Hartmann} L.,  {Calvet} N.,  {Gullbring} E.,    {D'Alessio} P.,  1998, \apj,
  495, 385

\bibitem[\protect\citeauthoryear{{Hawley}, {Gammie} \& {Balbus}}{{Hawley}
  et~al.}{1995}]{HawleyEtAl1995}
{Hawley} J.~F.,  {Gammie} C.~F.,    {Balbus} S.~A.,  1995, \apj, 440, {742
  (HGB)}

\bibitem[\protect\citeauthoryear{{Hawley} \& {Stone}}{{Hawley} \&
  {Stone}}{1998}]{HawleyAndStone1998}
{Hawley} J.~F.,  {Stone} J.~M.,  1998, \apj, 501, 758

\bibitem[\protect\citeauthoryear{{Hayashi}}{{Hayashi}}{1981}]{Hayashi1981}
{Hayashi} C.,  1981, Progress of Theoretical Physics Supplement, 70, 35

\bibitem[\protect\citeauthoryear{{Hirose}, {Krolik} \& {Blaes}}{{Hirose}
  et~al.}{2009}]{HiroseEtAl2009}
{Hirose} S.,  {Krolik} J.~H.,    {Blaes} O.,  2009, \apj, 691, 16

\bibitem[\protect\citeauthoryear{{Hirose}, {Krolik} \& {Stone}}{{Hirose}
  et~al.}{2006}]{HiroseEtAl2006}
{Hirose} S.,  {Krolik} J.~H.,    {Stone} J.~M.,  2006, \apj, 640, {901 (HKS)}

\bibitem[\protect\citeauthoryear{{Hirose} \& {Turner}}{{Hirose} \&
  {Turner}}{2011}]{HiroseAndTurner2011}
{Hirose} S.,  {Turner} N.~J.,  2011, \apjl, 732, L30+

\bibitem[\protect\citeauthoryear{{Ilgner} \& {Nelson}}{{Ilgner} \&
  {Nelson}}{2006}]{2006A&A...445..205I}
{Ilgner} M.,  {Nelson} R.~P.,  2006, \aap, 445, 205

\bibitem[\protect\citeauthoryear{{Johansen} \& {Klahr}}{{Johansen} \&
  {Klahr}}{2005}]{JohansenAndKlahr2005}
{Johansen} A.,  {Klahr} H.,  2005, \apj, 634, 1353

\bibitem[\protect\citeauthoryear{{King} \& {Pringle}}{{King} \&
  {Pringle}}{2010}]{KingAndPringle2010}
{King} A.~R.,  {Pringle} J.~E.,  2010, \mnras, 404, 1903

\bibitem[\protect\citeauthoryear{{King}, {Pringle} \& {Livio}}{{King}
  et~al.}{2007}]{KingEtAl07}
{King} A.~R.,  {Pringle} J.~E.,    {Livio} M.,  2007, \mnras, 376, 1740

\bibitem[\protect\citeauthoryear{{Kretke} \& {Lin}}{{Kretke} \&
  {Lin}}{2010}]{KretkeAndLin2010}
{Kretke} K.~A.,  {Lin} D.~N.~C.,  2010, \apj, 721, 1585

\bibitem[\protect\citeauthoryear{{Levermore} \& {Pomraning}}{{Levermore} \&
  {Pomraning}}{1981}]{LevermorePomraning1981}
{Levermore} C.~D.,  {Pomraning} G.~C.,  1981, \apj, 248, 321

\bibitem[\protect\citeauthoryear{{Miyoshi} \& {Kusano}}{{Miyoshi} \&
  {Kusano}}{2006}]{MiyoshiAndKusano2006}
{Miyoshi} T.,  {Kusano} K.,  2006, AGU Fall Meeting Abstracts, pp C275+

\bibitem[\protect\citeauthoryear{{Oppenheimer} \& {Dalgarno}}{{Oppenheimer} \&
  {Dalgarno}}{1974}]{OppenheimerDalgarno1974}
{Oppenheimer} M.,  {Dalgarno} A.,  1974, \apj, 192, 29

\bibitem[\protect\citeauthoryear{{Pollack}, {McKay} \&
  {Christofferson}}{{Pollack} et~al.}{1985}]{PollackEtAl1985}
{Pollack} J.~B.,  {McKay} C.~P.,    {Christofferson} B.~M.,  1985, \icarus, 64,
  471

\bibitem[\protect\citeauthoryear{{Sano} \& {Stone}}{{Sano} \&
  {Stone}}{2002}]{SanoAndStone2002}
{Sano} T.,  {Stone} J.~M.,  2002, \apj, 577, 534

\bibitem[\protect\citeauthoryear{{Simon}, {Armitage} \& {Beckwith}}{{Simon}
  et~al.}{2011}]{SimonEtAl2011}
{Simon} J.,  {Armitage} P.,    {Beckwith} K.,  2011, ArXiv e-prints

\bibitem[\protect\citeauthoryear{{Simon}, {Hawley} \& {Beckwith}}{{Simon}
  et~al.}{2011}]{SimonAndHawley2011}
{Simon} J.~B.,  {Hawley} J.~F.,    {Beckwith} K.,  2011, \apj, 730, 94

\bibitem[\protect\citeauthoryear{{Sorathia}, {Reynolds}, {Stone} \&
  {Beckwith}}{{Sorathia} et~al.}{2011}]{SorathiaEtAl2011}
{Sorathia} K.~A.,  {Reynolds} C.~S.,  {Stone} J.~M.,    {Beckwith} K.,  2011,
  ArXiv e-prints

\bibitem[\protect\citeauthoryear{{Stone} \& {Gardiner}}{{Stone} \&
  {Gardiner}}{2010}]{StoneEtAl2010}
{Stone} J.~M.,  {Gardiner} T.~A.,  2010, \apjs, 189, 142

\bibitem[\protect\citeauthoryear{{Stone}, {Hawley}, {Gammie} \&
  {Balbus}}{{Stone} et~al.}{1996}]{StoneEtAl1996}
{Stone} J.~M.,  {Hawley} J.~F.,  {Gammie} C.~F.,    {Balbus} S.~A.,  1996,
  \apj, 463, 656

\bibitem[\protect\citeauthoryear{{Turner}, {Carballido} \& {Sano}}{{Turner}
  et~al.}{2010}]{TurnerEtAl2010}
{Turner} N.~J.,  {Carballido} A.,    {Sano} T.,  2010, \apj, 708, 188

\bibitem[\protect\citeauthoryear{{Turner} \& {Sano}}{{Turner} \&
  {Sano}}{2008}]{TurnerAndSano2008}
{Turner} N.~J.,  {Sano} T.,  2008, \apjl, 679, L131

\bibitem[\protect\citeauthoryear{{Turner}, {Willacy}, {Bryden} \&
  {Yorke}}{{Turner} et~al.}{2006}]{TurnerEtAl2006}
{Turner} N.~J.,  {Willacy} K.,  {Bryden} G.,    {Yorke} H.~W.,  2006, \apj,
  639, 1218

\bibitem[\protect\citeauthoryear{{Umebayashi} \& {Nakano}}{{Umebayashi} \&
  {Nakano}}{2009}]{2009ApJ...690...69U}
{Umebayashi} T.,  {Nakano} T.,  2009, \apj, 690, 69

\end{thebibliography}
\end{document}